\documentclass[prb,showpacs,floatfix,twocolumn]{revtex4}
\usepackage{graphicx}
\usepackage{dcolumn}
\begin{document}
\newcommand*{\cm}{cm$^{-1}$\,}
\newcommand*{\nco}{Na$_x$CoO$_2$\,}
%
\title{Electron-boson mode coupling and the pseudogap of Na$_x$CoO$_2$ by infrared spectroscopy}
%
%
\author{D. Wu}
\affiliation{Beijing National Laboratory for Condensed Matter
Physics, Institute of Physics, Chinese Academy of Sciences,
Beijing 100080, China}
\author{J. L. Luo}
\affiliation{Beijing National Laboratory for Condensed Matter
Physics, Institute of Physics, Chinese Academy of Sciences,
Beijing 100080, China}
\author{N. L. Wang}
\altaffiliation{Corresponding author}\email{nlwang@aphy.iphy.ac.cn}%
\affiliation{Beijing National Laboratory for Condensed Matter
Physics, Institute of Physics, Chinese Academy of Sciences,
Beijing 100080, China}
\date{\today}
%
%
\begin{abstract}
We report the in-plane optical measurements on Na$_x$CoO$_2$ with
0.18$\leq$x$\leq$0.92. The crystal growth and characterization
were described in detail. The spectral weight increases
monotonically with decreasing x. For sample with the lowest Na
content x$\sim$0.18, the optical conductivity was strongly
suppressed below 2200 $cm^{-1}$ at low temperature. The
suppression becomes weaker with increasing Na content, and
disappears in charge-ordered sample with x$\sim$0.5. At higher Na
contents, similar suppression appears again but locates at higher
energy scale near 3300 $cm^{-1}$. Our analysis indicates that
those spectral features are dominated by a combination of
electrons coupling to a bosonic mode and a pseudogap-like
phenomenon. We suggest that the pseudogap-like phenomenon is
purely a band structure effect. The infrared activated phonon
modes were discussed in relation with the structural geometries.
\end{abstract}

\maketitle

\section{INTRODUCTION}

The discovery of superconductivity at 4 K in hydrated sodium
cobaltate\cite{Takada} has attracted much attention in
Na$_x$CoO$_2$ material. A rich phase diagram has been revealed for
Na$_x$CoO$_2$ with a change of Na content x. A spin ordered phase
is found for x$>$3/4. With decreasing Na contents, the material
becomes a "Curie-Weiss metal" for x near 2/3, then a
charge-ordered insulator with x$\sim$ 1/2, and finally a
paramagnetic metal with x$\sim$ 1/3.\cite{Foo} Superconductivity
occurs when sufficient water is intercalated between the CoO$_2$
layers for x near 1/3. The cobaltate provides a model system for
studying the physics of correlated electrons in a 2D triangular
lattice. It is also widely expected that the study of \nco system
may shed new light on high-T$_c$ superconductivity in cuprates.

Various experimental techniques have been employed to explore the
properties of \nco. Among others, the optical spectroscopy yields
important information about electronic structure and charge
dynamics of the system. Several groups
\cite{Lupi,Wang1,Bernhard,Caimi,Hwang1,Wang2} have performed
optical studies on \nco, but those work were almost limited to
compounds with high Na concentration (x$\geq$0.5). The only work
\cite{Hwang1} that contains infrared data of a x=0.25 compound
revealed an insulating behavior of the sample, and its spectral
feature is almost identical to that seen on charge-ordered
compound of x=0.5.\cite{Hwang1,Wang2} This result is in
contradiction to the expected metallic response at this doping
level. Therefore, the so-called paramagnetic metals for x near
1/3, a doping region where superconductivity occurs after
hydrated, remain unexplored by optical spectroscopy.

On the other hand, the reported optical data show apparent
controversies, particularly with respect to the low-$\omega$
charge dynamics. Lupi et al.\cite{Lupi} measured the optical
spectra of a Na$_{0.57}$CoO$_2$ single crystal and reported an
anomalous Drude behavior. The optical scattering rate,
1/$\tau(\omega)$, extracted from the extended Drude model shows a
$\omega^{3/2}$ dependence. Caimi et al.\cite{Caimi} studied a
Na$_{0.7}$CoO$_2$ crystal and found a roughly $\omega$-linear
dependence of the 1/$\tau(\omega)$ at low $\omega$. By contrast,
the recent work by Hwang et al.\cite{Hwang1} indicated a
$\omega$-linear dependence of 1/$\tau(\omega)$ at high $\omega$,
while the low-$\omega$ scattering is dominated by a bosonic mode
with an onset frequency of scattering near 600 \cm. The spectral
feature is very much like that in the cuprates. In this work, we
present the in-plane optical measurements at different
temperatures on Na$_x$CoO$_2$ with different x. We show that in
metallic samples, the optical spectra were dominated by a
combination of the coupling effect of electrons with a Boson mode
and a pseudogap-like phenomenon. We provide a detailed analysis
for the origin of the pseudogap-like feature.

\section{Crystal growth and characterizations}

High-quality Na$_x$CoO$_2$ single crystals were grown from a
floating zone optical image furnace. Powders of Na$_2$CO$_3$ and
Co$_3$O$_4$ with nominal chemical ratios of x=0.75$\sim$1.0 were
well mixed, ground, heated at 750$^{o}C$ for 24 hours twice with
intermediate grinding, followed by calcination at
820$\sim$850$^{o}C$ for another 24 hours. From X-ray diffraction
measurement, the calcined powders were already in a single phase.
Then, the calcined powders were pressed into a rod with a size
roughly 8 mm $\phi$$\times$10 cm under a pressure of 70 MPa, and
annealed at 850$^{o}C$ for one day in flowing oxygen to form feed
rod. A part of the rod ($\sim$2 $cm$) was cut and used as the
seed. The feed rod was pre-melted in the floating zone furnace
with a feeding speed of 30 mm$/$h. Then the pre-melted rod was
used as a feed and re-melted with a much slower feeding speed
2$\sim$4 mm$/$h under 5 atm oxygen pressure. During the procedure,
the feed rod and seed were rotated in contrary directions at a
speed of 20 rpm to insure a uniform melting zone. The obtained
single crystal ingot was characterized by x-ray diffraction,
electrical transport and magnetic measurement. It is found that
the top 2 $cm$ part of the ingot has higher quality than the rest
part. So for every ingot, the top 2 cm was cut-off and the rest
part was re-melted again. The induction coupled plasma (ICP)
analysis was used to determine the Na:Co ratios. The sodium
concentrations in the range of 0.75$\sim$0.95 were obtained for
the as-grown crystals. For getting crystals with low sodium
contents, the obtained as-grown single crystals of
Na$_{0.85}$CoO$_2$ were immersed in solution of 60 ml
bromine/CH$_3$CN with very different bromine concentrations for
over one month. Eventually, we obtained a series of Na$_x$CoO$_2$
single crystals with reduced Na concentrations x=0.48, 0.36, 0.32,
0.18 by ICP measurement.

Fig. 1 (a) shows the (00l) x-ray diffraction patterns of cleaved
Na$x$Co$O_2$ single crystals with x=0.18, 0.32, 0.36, 0.48, 0.85,
and 0.92. The obtained c-axis lattice constants show a monotonous
increase for decreasing the sodium contents, as displayed in 1
(b). For three low Na content samples, the weak lines at
2$\theta$=12.8, 25.7, 36.7 are originated from water intercalation
after exposed to atmosphere for a long time.\cite{Lin} Samples
with higher Na contents x$>$0.48 are very stable, and no water
absorption could be detected. An as-grown Na$_0.85$CoO$_2$ crystal
ingot is shown in 1 (c). The ab-plane can be easily cleaved from
the ingot due to layered structure of the compound.

\begin{figure}
\includegraphics[width=0.9\linewidth]{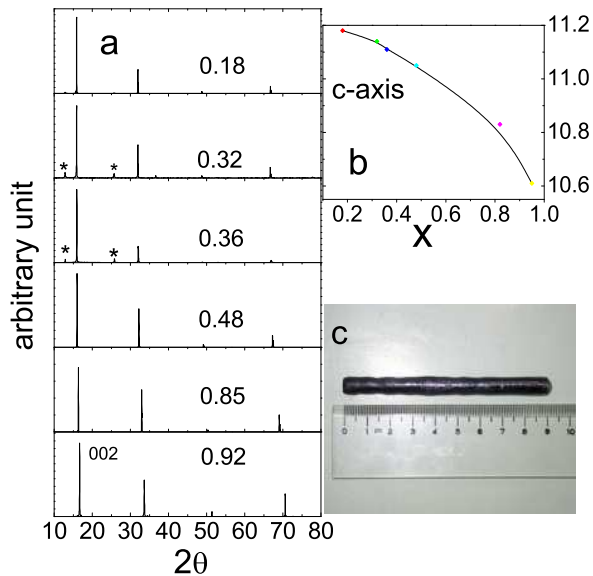}
\caption{\label{fig:x-ray}{(Color online) (a) x-ray diffraction
patterns of (00l) planes for different compositions. The weak
lines marked with $*$ were due to water intercalations when
exposed to atmosphere for over 10 days. (b) the c-axis lattice
constant $vs$ sodium content. (c) a typical ingot of as-grown
Na$_0.85$CoO$_2$.}}
\end{figure}

Fig. 2 shows the in-plane resistivity $\rho$ for several
compositions measured by a four-probe low frequency ac method. For
the as-grown x$=$0.85 sample, it shows a metallic behavior above
18 K but exhibits an upturn at low temperature because of the
formation of the A-type long-range anti-ferromagnetic
ordering\cite{Luo,Boothroyd}. A weak upturn feature is also seen
for x=0.92 sample. For x=0.48 crystal, a sharp metal-insulator
transition occurs at 48 K which is due to the well-known
charge-ordering formation. The two other samples with lower Na
contents of x=0.32 and 0.18 show metallic behavior in the whole
temperature range. For the measured samples, the in-plane
resistivity decreases monotonously with decreasing Na content.
Since the x=0 compound is a half-filled system (with a
t$_{2g}^5$e$_g^0$ (Co$^{+4}$) electron configuration), and is
anticipated to be a Mott insulator within the strong
electron-correlation picture. One would expect to observe a
transition from metallic to insulating transport for x approaching
zero. However, down to the lowest Na content x=0.18, there is no
indication for such a transition. As the compound with lower Na
content was not obtained, it is not clear whether the x=0 compound
is really a Mott insulator or a good metal with half-filled band.

\begin{figure}
\includegraphics[width=8.5cm]{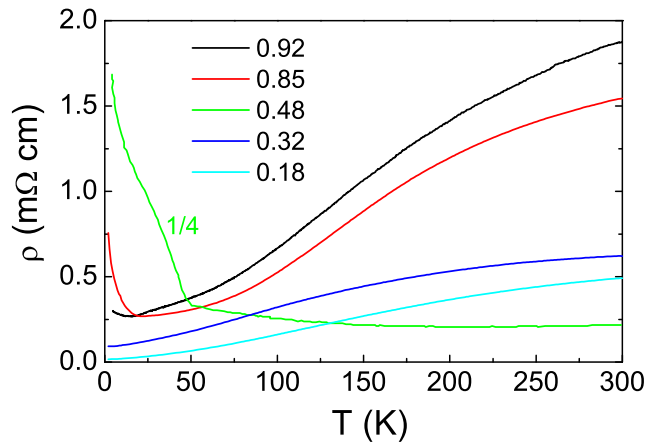}
\caption{\label{fig:r-T}{(Color online) The in-plane resistivity
$\rho$ $vs$ temperature for several \nco samples. Note the factor
used to scale the curves for the x=0.48 sample.}}
\end{figure}

\section{Optical properties}

\subsection{Evolution of optical spectra with doping}

The near-normal incident reflectance spectra were measured on the
freshly cleaved surface using a Bruker 66v/s spectrometer in the
frequency range from 40 \cm to 28000 \cm, as described in our
earlier report.\cite{Wang1,Wang2} Standard Kramers-Kronig
transformations were employed to derive the frequency-dependent
conductivity spectra. R($\omega$) was extrapolated to zero
frequency by using a standard Hagen-Rubens behavior and a constant
extrapolation to 400000 \cm followed by a $\omega^{-4}$ behavior
in high energy side.

\begin{figure}
\includegraphics[width=8.5cm]{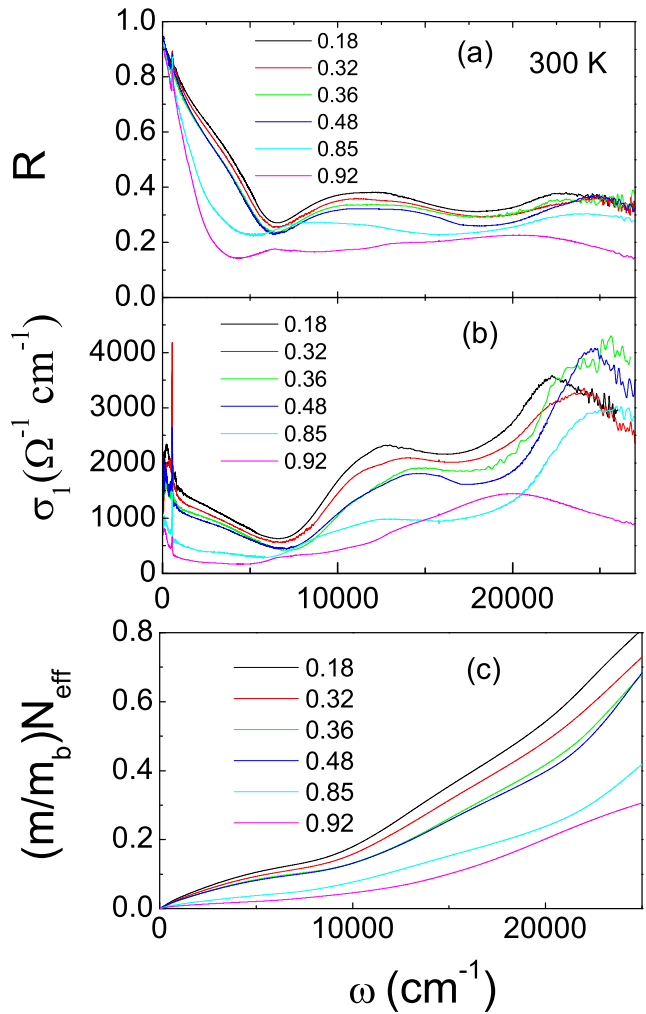}
\caption{\label{fig:R-S-N-300k}{(Color online) Fig.3 (a) and (b)
show the doping evolution of in-plane reflectance and conductivity
of \nco at room temperature over broad frequencies. Panel (c)
displays the effective carrier density of Co ion for different
sodium contents.}}
\end{figure}

Fig. 3 (a) and (b) show the room temperature in-plane reflectance
and conductivity spectra for Na$_x$CoO$_2$ with x=0.18$\sim$0.92
over broad frequencies. We found that the compound with the
highest Na content has the lowest reflectance values and edge
frequency, as a result, it has the lowest spectral weight in
optical conductivity curve. According to the partial sum-rule, the
area under the conductivity curve below a certain frequency
corresponds to the effective carrier density below that frequency.
The effective density of carriers per Co ion contributed to
conductivity below $\omega$ is
$(m/m_b)N_{eff}(\omega)=(2mV_{cell}/{\pi}e^2N)\int_0^{\omega}\sigma(\omega')d\omega'$,
where m is the free-electron mass, m$_b$ the averaged
high-frequency optical or band mass, $V_{cell}$ a unit cell
volume, N the number of Co ions per unit volume. Fig. 1(c)
displays $(m/m_b)N_{eff}$ as a function of frequency for those
samples. The result is in good agreement with earlier studies on
limited doping levels.\cite{Wang2,Hwang1} Obviously, the compound
with high Na content has low conducting carrier density. Since the
x=1 compound has a t$_{2g}^6$e$_g^0$ (low spin state of Co$^{3+}$)
electron configuration and is expected to be a band insulator with
completely filled t$_{2g}$ band and empty e$_g$ band, the
gradually vanishing spectral weight for high Na content samples
confirms such a band insulator for x=1 compound. With reducing the
sodium content from 0.92 to 0.18, the spectral weight increases
monotonously, suggesting a continuous increase of the effective
carriers density. Many people considered the \nco system in the
superconducting region (x$\sim$0.3) as being evolved from the x=0
compound by adding the system electrons, i.e. a doped Mott
insulator. The optical data here indicate that at least down to
x=0.18, the effective carrier density still increases with
decreasing Na content, favoring a doped band insulator picture
rather than a doped Mott insulator. We noticed that spectral
increase for x below 0.5 becomes less as fast as it does for
x$>$0.5, however, there is no indication for a decrease of the
spectral weight with further reducing Na content.

Another important feature is that all samples have two apparent
interband transitions near 1.6 eV and 3.1 eV, except for the
x=0.92 compound which shows roughly a single broad interband
transition centered around 2.5 eV. This energy scale is in good
agreement with the simple LDA calculation.\cite{Singh} A naive
interpretation for the two peak structure is that the low-energy
peak is originated from the interband transition between the
occupied Co t$_{2g}$ to unoccupied Co e$_g$ bands, while the
high-energy one is due to the transition from occupied O $_{2p}$
to Co e$_g$ bands.\cite{Johannes} Unfortunately, the observed
energy scales are much lower than the LDA band structure
calculations. The interpretation also has the difficulty to
explain the broad single interband transition peak observed for
x=0.92 sample with the energy scale now in good agreement with the
simple LDA calculations for transition between t$_{2g}$ and e$_g$
bands. In our previous work on Na$_{0.7}$CoO$_2$,\cite{Wang1} we
interpreted both peaks near 1.6 eV and 3.1 eV as interband
transitions from occupied t$_{2g}$ to empty e$_g$ bands by
invoking an exchange splitting. The present results add support
for this interpretation. Note that, as x approaches 1 (band
insulator), the t$_{2g}$ band tends to be fully filled and e$_g$
band completely empty, the exchange splitting would disappear. In
this case, a single interband transition from t$_{2g}$ to e$_g$
with energy scale close to the simple LDA band structure
calculation is expected. The x=0.92 sample indeed follows this
expectation.

\subsection{Temperature-dependent optical response: mode coupling and gap-like structure}

The temperature-dependent $R(\omega)$ spectra are shown in figure
4. At room temperature, all samples show metallic behavior,
indicated by the high reflectance at low frequency and a gradually
decreasing behavior up to a reflectance edge. For each sample, the
spectra of different temperatures cross at certain frequencies,
indicating a spectral weight transfer between high and low energy
region. The temperature dependences of the optical responses are
also metallic except for those samples with high sodium contents
at low temperature. For the x=0.85 and 0.92 samples, the
reflectances below 250 cm$^{-1}$ are clearly depressed at T=10 K,
leading to a suppression of conductivity spectra. The optical
result is consistent with the dc resistivity measurement showing
an upturn below 20 K due to AF ordering forming\cite{Luo}. The
other one is the insulating charge-ordering sample x=0.48. Its
spectra are very similar to our reported data on x=0.50
compound\cite{Wang2} showing characteristic spectral features: an
opening of gap below 125 cm$^{-1}$ together with the development
of an electronic resonance peak near 800 cm$^{-1}$, as seen
clearly in Fig. 5.

The upper panels of figure 5 show the T-dependent
$\sigma_1(\omega)$ spectra obtained from their reflectances by
Kramers-Kroning transformations. The data on x=0.85 and 0.92
compounds are similar to our previous data on x=0.7
compound,\cite{Wang1} except that the present samples have lower
$\sigma_1(\omega)$ values. The compounds with lower Na content
0.18$\leq$x$\leq$0.36 show metallic response and have apparently
higher conductivity values. In the very far-infrared region, an
unusual drop in $\sigma_1(\omega)$ is commonly observed for those
samples, leading to a peak at finite energy ($\sim$200 \cm). This
feature was addressed in our earlier work.\cite{Wang1} The most
striking observation here is that for x=0.18, 0.32 and 0.36
samples, the $\sigma_1(\omega)$ spectra show distinct suppressions
at low T below about 2200 \cm. Such suppression was usually taken
as a signature of a pseudogap (PG) in $\sigma_1(\omega)$ spectra.
We notice that this suppression structure is strongest for x=0.18
sample, becomes weaker with increasing Na content, and disappears
completely when approaching the charge-ordered insulating phase
near x=0.5, for which instead of a suppression we observe an
pronounced electronic resonance peak near 800 \cm. However, at
higher Na contents, the suppression feature appears again, but its
peak position occurs at higher energy near 3300 \cm, which was
previously labelled as the $\gamma$-peak.\cite{Wang1}.

\begin{figure}
\includegraphics[width=8cm]{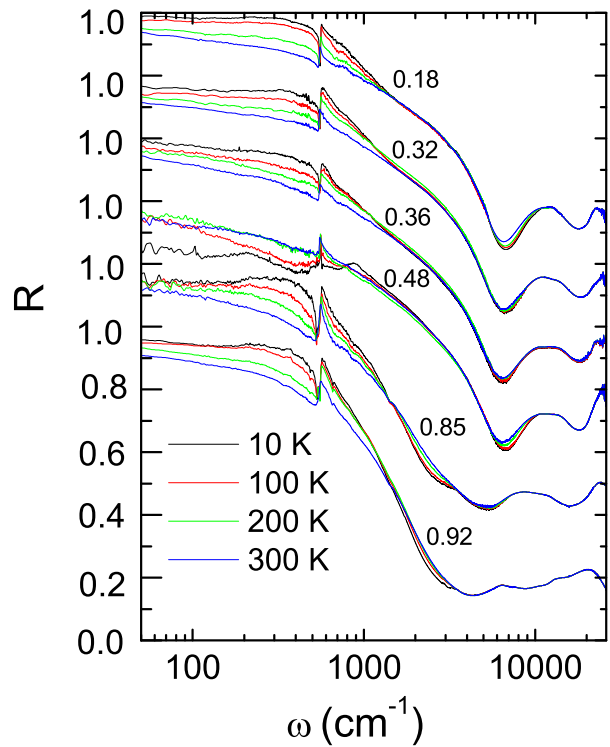}
\caption{\label{fig:R-S-N}{(Color online) Fig.4 shows the
T-dependant in-plane reflectance of \nco system. All compounds
show metallic response with decreasing temperature except for the
charge ordering insulating sample x=0.48, and two high sodium
content samples x=0.85 and x=0.92 where the low-T reflectance was
depressed below 250 \cm due to the formation of the A-type AF
magnetic ordering.}}
\end{figure}

Although the conductivity displays a gap-like suppression at low
temperature, an integration of the conductivity up to 5000
cm$^{-1}$ does not show a decrease of the spectral weight within
our experimental uncertainty. The depression of the conductivity
in the mid-infrared region is balanced out by the increase of
conductivity at low frequency. Despite the presence of a peak at
finite frequency, the low-frequency Drude-like conductivity shows
a rapid narrowing with decreasing temperature. We emphasize that
such spectral change resembles significantly to the hope-doped
cuprates in the PG state, where no detectable SW loss was found in
the ab-plane conductivity as well.\cite{Syro} Understanding the
low-frequency dynamics, including the rapid narrowing behavior of
the low-frequency component and the PG-like suppression readily
seen at low temperature, is our major focus in this paper.

\begin{figure*}
\centerline{\includegraphics[width=7.0in]{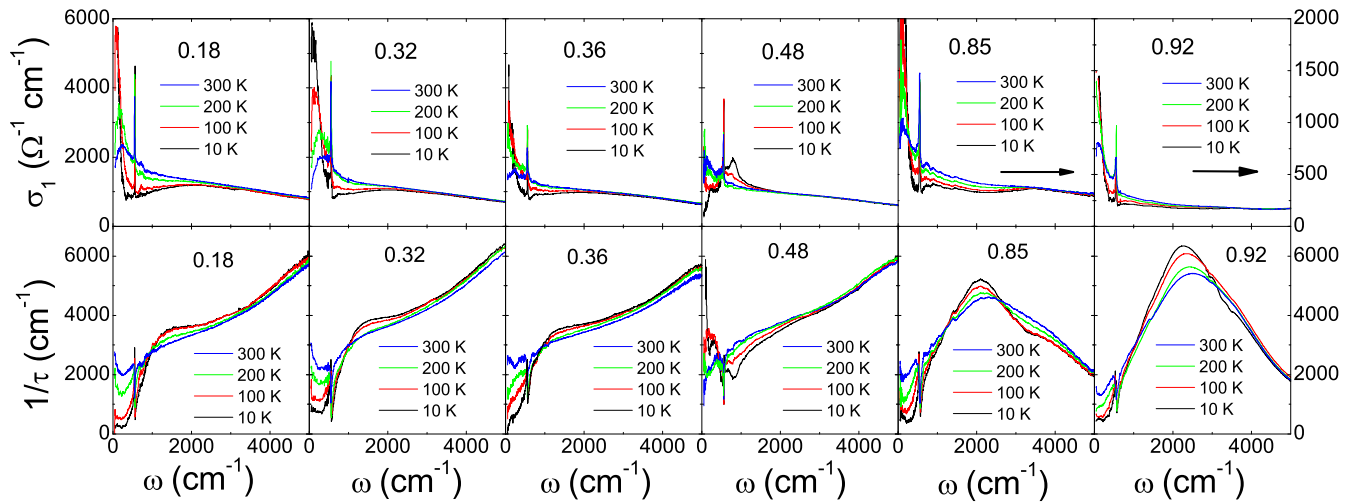}}%
\caption{(Color online) The T-dependence of the
conductivity and scattering rate spectra for \nco with different x.}%
\label{fig5}
\end{figure*}

It is well known that, for the case of metallic conduction in the
ab-plane, the signature of a PG is best resolved in the spectrum
of the $\omega$-dependent scattering rate.\cite{Timusk} The bottom
panels of Fig. 5 show the 1/$\tau(\omega)$ spectra for above six
samples obtained from the extended Drude model\cite{Puchkov}
1/$\tau(\omega)$= $(\omega_p^2 / 4\pi)$Re(1/$(\sigma(\omega))$,
where $\omega_p$ is the overall plasma frequency and can be
obtained by summarizing the $\sigma_1(\omega)$ up to the
reflectance edge frequency. Related to the suppression in
$\sigma_1(\omega)$, we observe a strong suppression feature as
well in 1/$\tau(\omega)$ at frequencies near 1300 \cm for
x=0.18$\sim$0.36 samples. Compared with data at high temperature,
the low-T spectra show weak "overshoot" behaviors just above the
suppression frequencies. We notice again that the spectral feature
in 1/$\tau(\omega)$ is similar to that of high-Tc
cuprates\cite{Timusk,Puchkov} and other materials with partially
gaped Fermi surface (FS) like, for example, an antiferromagnet Cr
for which a spin-density-wave gap opens in parts of the
FS\cite{Basov1}. Note that the suppression in 1/$\tau(\omega)$
appears at lower frequency than that in $\sigma_1(\omega)$
spectra. This is a generic behavior for PG, and was discussed in
our recent work on cuprate superconductors.\cite{Wang3} Another
notable observation is that, different from samples with low Na
contents, the x=0.85 and 0.92 samples display prominent peaks near
2000 \cm in 1/$\tau(\omega)$ spectra. This is because the free
carrier contribution to the high-$\omega$ spectral weight is very
low, which makes the gap or interband transition structures more
pronounced. Here we show the 1/$\tau(\omega)$ spectra for all
samples obtained by the same method, but we should bear in mind
that the non-monotonous spectral shape implies a breakdown of the
extended Drude mode for application to these compounds. We found
again that the peak position in 1/$\tau(\omega)$ is much lower
than the suppression energy in $\sigma_1(\omega)$.

It should be pointed out that there is no simple and direct
connection between the quantity $\sigma_1(\omega)$ or
1/$\tau(\omega)$ and the density of states (DOS). An exact
relationship depends on model. Simply from the suppressions in
$\sigma_1(\omega)$ or 1/$\tau(\omega)$ spectra, one cannot
conclude the presence of a true gap structure in the DOS, because
very similar spectral change can also be caused by a strong
coupling effect of electrons with a bosonic mode. In the later
case, the optical spectra would display strongly enhanced
absorption above the mode energy, which leads to a sharp increase
in the 1/$\tau(\omega)$. This result can be verified easily from
famous Allen's formula for the case of electron-phonon coupling at
zero temperature,\cite{Allen} ${1/\tau(\omega)}$=
   ${(2\pi/\omega})\int_0^{\omega}d\Omega(\omega-\Omega)\alpha^2F(\Omega)$,
by assuming a peak-like phonon spectral function of
$\alpha^2F(\Omega)$. The strong coupling effect of electrons with
a collective mode was intensively studied in high-T$_c$
cuprates.\cite{Carbotte,Abanov,Tu,Wang4,Hwang2} The same effect
was also suggested in earlier optical study on Na$_{0.75}$CoO$_2$
by Hwang et al.\cite{Hwang1}. From the analysis of the real and
imaginary parts of the optical scattering rate (or the optical
self-energy), they extracted an underlying bosonic mode with
frequency near 600 \cm. Recent ARPES experiments on Na$_x$CoO$_2$
also reveal a mode coupling effect near 70-80 meV based on a
self-energy analysis.\cite{Qian} In the present work, we can
identify from figure 5 that all samples, except the charge-ordered
insulating one, exhibit a sharp increase structure at
$\omega\sim$500-600 \cm in 1/$\tau(\omega)$ spectra. The feature
is strongest at lowest T, but becomes less prominent with
increasing T. Furthermore, there is no shift in energy scale for
samples with different x. Naturally, we can conclude that the mode
is indeed present in all those samples and results in similar
features in their optical spectra. We noticed that the phonons are
in the right frequency range for the underlying mode.

A crucial issue is can we distinguish the effect resulted by a PG
or a bosonic mode from those optical spectra? In other words,
whether the above mentioned characteristic features in optical
spectra are contributed by one of the two phenomena or by a
combination of both that appear together? A critical analysis of
the spectral shape provides a clue to this question. The key point
is that the electron-boson coupling can only result in an increase
in 1/$\tau(\omega)$ spectrum, it can never cause an overshoot or
peak in 1/$\tau(\omega)$. Therefore, the observation of the
overshoot behavior must have a different origin. This statement is
further strengthened by the result of inverted bosonic spectral
function, as we shall explain below.

A simple way to extract the electron-Boson spectral function is to
use the formula by Marsiglio et al.\cite{Marsiglio}, $W(\omega)$=
$(1/2\pi)$$d^2[\omega/\tau(\omega)]/d\omega^2$. Since the second
derivative is used, heavy smoothing of the experimental data has
to be done. In the present work, we first remove a phonon near 560
\cm in 1/$\tau(\omega)$, then fit the spectra with high-order
polynomial. As all metallic samples have similar spectra, we only
show the results for x=0.18 sample. Fig. 6(a) shows the spectral
function obtained from the above formula. The inset shows the
1/$\tau(\omega)$ at 10 K together with the fitting curve with
phonon removed. The most prominent feature in W($\omega$) is a
large maximum near 530 \cm and a deep negative minimum at 1300
\cm. Again the spectral function resembles significantly to the
cuprate superconductors.\cite{Carbotte,Abanov,Tu,Wang4} The large
maximum corresponds to the sharp increase in 1/$\tau(\omega)$,
while the negative minimum is linked with the weak overshoot.
Since the electron-boson spectral density cannot be negative, the
negative dip in W($\omega$) could not originate from the
electron-boson coupling.

In fact, earlier studies on a BCS superconductor have revealed a
peak structure in scattering rate spectrum due to the presence of
superconducting gap in the density of states
(DOS).\cite{Basov1,Wang3} Obviously, if we perform second
derivative, we would get a negative minimum in W($\omega$).
Therefore, a gap (or pseudogap) in DOS can cause such negative
structure in W$(\omega)$. Taking account of those different
effects, we can tentatively conclude that both the electron-boson
coupling effect and the pseudogap-like phenomenon are likely
present in \nco.

It would be most desirable to quantitatively analyze the low-T
1/$\tau(\omega)$ for the case of electron-Boson coupling in the
presence of a gap in DOS. Recently, several groups made such
attempts to low-T 1/$\tau(\omega)$ of high-T$_c$
cuprates.\cite{Dordevic,Hwang3} In fact, Allen generalized the
electron-phonon coupling formula at T=0 K for a BCS superconductor
as, ${1/\tau(\omega)}$=
${(2\pi/\omega})\int_0^{\omega-2\Delta}d\Omega(\omega-\Omega)\alpha^2F(\Omega)E[\sqrt{1-4\Delta^2/(\omega-\Omega)^2}]$,
where E(x) is the second elliptic function.\cite{Allen} Dordevic
et al.\cite{Dordevic} employed this formula to the data of YBCO,
and found that it reproduced fairly well the low-T
1/$\tau(\omega)$ with the use of a finite value of the gap and a
positive bosonic spectral function. More recently, Hwang et
al.\cite{Hwang3} used a more generalized formula to the data of
ortho-II YBCO. In their formula, the shape of the gap can be
modelled flexibly. Although we found that those approaches can
also reproduce well the low-T 1/$\tau(\omega)$ of \nco, there is a
strong reason against the application of those approaches here,
for those formulae were derived based on the assumption that a
superconducting gap opens below T$_c$. Furthermore, the impurity
elastic scattering was ignored. Therefore, those approaches are
valid only for superconductors at T$<<$T$_c$ in the clean limit.
While in the case of \nco, the samples we studied are not
superconducting at all.

According to the band structure calculation on \nco\cite{Singh},
two bands, an a$_{1g}$ and one of the e$_g^\prime$ bands, from Co
t$_{2g}$ manifold cross the Fermi level (E$_F$). They form a large
hole Fermi surface (FS) centered at $\Gamma$ (by a$_{1g}$) and six
small Fermi pockets near the zone boundary (by e$_g^\prime$). The
large FS was confirmed by the ARPES measurements, however, the six
pockets were not observed\cite{Hasan,Yang}. It deserves to point
out that, unlike cuprate superconductors where partial states are
really removed in FS near "hot spots", or a d-wave gap really
exists at low T, ARPES experiments on all available metallic \nco
indicated no sign of gap opening on any part of the large
FS\cite{Hasan,Yang,Yang2}. This makes the gap-like suppression
more puzzling in \nco than in cuprates. One cannot conclude that
the gap-like feature is related to the large FS.

One might link the PG feature to the missed small FS pockets.
ARPES experiments indicated that the e$_g^\prime$ pockets near K
points sink down $\sim$ 200 meV below E$_F$, and have almost no
doping or T dependence.\cite{Yang2} Band structure calculation
also indicates that the pocket band may not touch E$_F$, if a
proper on-site Coulomb repulsion U is considered.\cite{Zhang} At
the first glance, this energy scale seen by ARPES is very close to
the suppression energy in 1/$\tau(\omega)$. However, because there
exists no other band above E$_F$ at the same momentum in the
reciprocal space where the pocket band sinks, no allowed optical
interband transition from occupied e$_g^\prime$ bands to others
just above E$_F$ exists. This rules out the possibility.

\begin{figure}
\includegraphics[width=8cm]{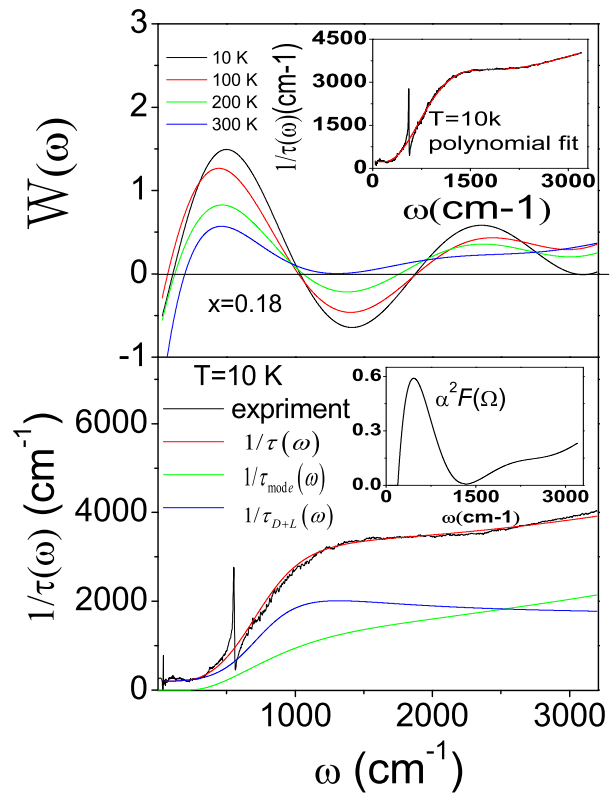}
\caption{\label{fig:W-S}{(Color online) (a) The bosonic spectral
function $W(\omega)$. Inset shows the $1/\tau(\omega)$ at 10 K
together with a 40-term poly-nomial fit. (b) The comparison of
experimental and calculated $1/\tau(\omega)$ spectra. Inset shows
the positive bosonic spectral function used for calculation.}}
\end{figure}

We noticed that the interband transition from occupied
e$_g^\prime$ to unoccupied a$_{1g}$ bands in other momentums
within t$_{2g}$ manifold are possible. This is actually one of
explanations we suggested for the origin of the $\gamma$ peak in
ref.\cite{Wang1}. It appears to be the most likely interpretation
for the origin. Indeed, we found that the optical scattering rate
can be modelled by a formula containing a mid-infrared component,
which could be naturally ascribed to such interband transition. To
illustrate, we assume that the 1/$\tau(\omega)$ was contributed by
a bosonic mode together with two other components (Drude+Lorentz):
${1/\tau(\omega)}$=${1/\tau_{mode}}$+${1/\tau_{(D+L)}}$ where
${1/\tau_{mode}}$
=${(2\pi/\omega})\int_0^{\omega}d\Omega(\omega-\Omega)\alpha^2F(\Omega)$,
and ${1/\tau_{(D+L)}}$= $(\omega_p^2/4\pi)$Re(1/$\sigma_{(D+L)}$)
with $\sigma_{(D+L)}$=$(\omega_{pD}^2/4\pi)[i/(\omega+i\tau_D)]$ +
$(\omega_{pL}^2/4\pi)[-i\omega/((\omega_0^2-\omega^2)-i\omega\tau_L)]$.
Indeed, we found that this expression can reproduce the data well
with the following parameters $\omega_p$=16500 \cm,
$\omega_{pD}$=11180 \cm, $\tau_D$=90 \cm, $\omega_{pL}$=26450 \cm,
$\omega_0$=2200 \cm, $\tau_L$=6000 \cm, and a bosonic mode showing
in the inset of Fig. 3(b). We emphasize that, in this approach,
the bosonic mode is still necessary in order to give a continuous
increase of scattering rate at high $\omega$. The above analysis
suggest that the PG feature could be purely a band structure
effect. Recent calculation for $\sigma(\omega)$ based on band
structure can indeed give a weak peak at low
$\omega$\cite{Johannes}. Based on those results, we can conclude
that both the mode coupling and interband transition contribute to
the low temperature spectral feature in \nco system. Furthermore,
the band structure may experience a significant change when
crossing the charge-ordered phase near x=0.5, since the energy
scale in the suppression is substantially different.

\begin{figure}
\includegraphics[width=8.5cm]{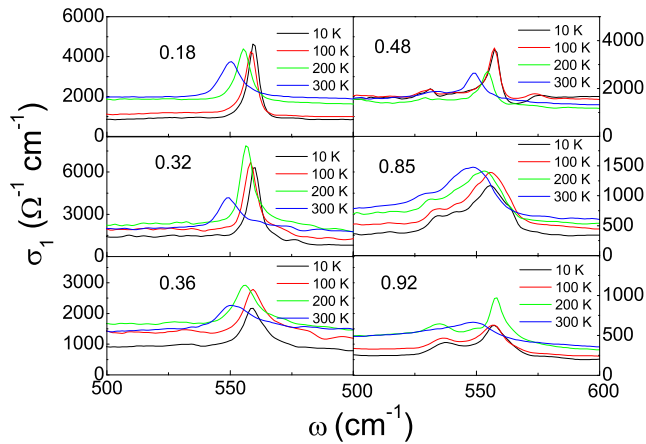}
\caption{(Color online)\label{fig:phonon} {(Color online) Fig. 7
shows the in-plane phonon spectra for \nco system. At room
temperature, all six samples have a phonon mode near 550
$cm^{-1}$. For $x\geq$0.48, another phonon mode near 530$cm^{-1}$
is observed.}}
\end{figure}

\subsection{Infrared phonon spectra}

Fig. 7 shows the conductivity spectra in an expanded region where
in-plane phonon modes are seen. For the three low sodium content
samples (left panels), there is only one infrared-active phonon
mode near 550 \cm at room temperature. With decreasing temperature
down to 10K, the phonon mode narrows and shifts slightly to the
high energy. For other three samples with higher sodium contents
$x\geq$0.48 (right panels), additional phonon mode near 530 \cm
can be seen at room temperature, besides the one at 550 \cm. The
mode near 530 $cm^{-1}$ becomes more eminent with increasing
sodium content, and also shifts to higher frequency with
decreasing temperature. In Na$_x$CoO$_2$, Na ions have two
possible sites: one is underneath the Co atom (being referred to
as A site), the other underneath the center of the triangle formed
by three neighboring Co atoms (B site), leading to two kinds of
structural geometries. The two Na sites cannot be simultaneously
occupied in the same unit cell because of impossibly close Na1-Na2
distances. According to the symmetry analysis, four infrared
active phonons $\Gamma_{IR}=2A_{2u}+2E_{1u}$ (two for
out-of-plane, two for in-plane vibrations) are predicted for
structure with only one Na site occupation.\cite{Li} The
respective mode frequencies are different for the two different
structural geometries. According to the calculation by Li et
al.\cite{Li}, the energy of phonon mode for geometry B is a little
bit higher than that of geometry A. The observation of two
in-plane phonon modes at room temperature, that are all close to
the frequency of hard E$_{1u}$ mode, may indicate that both
structural geometries are present in the sample. We also found
that for the x=0.85 sample, two additional peak/shoulder features
could be resolved at higher energy side of the two phonon modes at
low temperature. The result is similar to our earlier study on
x=0.7 sample.\cite{Wang1} The appearance of additional phonon
modes might be due to a structural change at low temperature.

\section{Summary}

In summary, the single crystals of Na$_x$CoO$_2$ were prepared
using floating zone method and characterized by x-ray diffraction
and dc resistivity measurements. The in-plane optical measurements
at different T on different x were measured. The vanishing
spectral weight for x approaching 1 confirms the expected band
insulator.  The spectral weight is found to increase monotonically
with decreasing Na content down to the lowest achieved x=0.18. The
result suggests against a doped Mott insulator for the system even
in the region where the compounds become superconducting after
hydrated. For sample with the lowest Na content x$\sim$0.18, the
optical conductivity was strongly suppressed below 2200 \cm at low
T. The suppression becomes weaker with increasing Na content, and
disappears when approaching charge-ordered phase with x$\sim$0.5,
where an insulating gap below 125 \cm and an electronic resonance
near 800 \cm develop at low temperature. After crossing the
charge-ordering phase, similar suppression appears again but at
higher energy scale near 3300 \cm. An partial sum rule analysis
indicates that there is no apparent loss of spectral weight. Then,
the low-frequency Drude-like component shows a rapid narrowing
behavior, correspondingly a strong depression is seen in the
optical scattering rate spectrum. The spectral features resemble
significantly to the hole-doped cuprates at low temperature. Our
analysis demonstrated that the optical spectra in those metallic
samples were caused by a combination of the coupling effect of
electrons with a Bosonic mode and a PG-like phenomenon. We suggest
that the PG-like phenomenon is purely a band structure effect, and
the interband transition within t$_{2g}$ manifold is responsible
for the PG-like feature. The in-plane phonon modes were displayed
and their evolution with x and temperature was discussed in
relation with the structural geometries.

\begin{acknowledgments}
We acknowledge discussions with H. Ding, T. Xiang, and G. M.
Zhang, and Y. C. Ma. This work was supported by National Science
Foundation of China, the Knowledge Innovation Project of Chinese
Academy of Sciences, and the Ministry of Science and Technology of
China£¨973 project No:2006CB601002).
\end{acknowledgments}

\end{document}